\documentclass[useAMS,usenatbib]{mn2e}
\usepackage{graphicx}
\usepackage{txfonts}
\usepackage{natbib}
\newcommand{\gro}{GRO~J1750$-$27}
\newcommand{\integral}{\emph{INTEGRAL}}
\newcommand{\swift}{\emph{Swift}}
\newcommand{\swiftbat}{\emph{Swift}/BAT}

\newcommand{\osa}{{\sevensize OSA7}}
\newcommand{\correction}{}

\title[Accretion powered spin-up of \gro] {The accretion powered spin-up of \gro}\author[S.E. Shaw et al.]{S.E.~Shaw$^1$, A.B.~Hill$^1$, E.~Kuulkers$^2$, S.~Brandt$^3$, J.~Chenevez$^3$, P.~Kretschmar$^2$\\$^1$School of Physics and Astronomy, University of Southampton, Highfield, Southampton, SO17 1BJ, UK.\\
$^2$ISOC, ESA/ESAC, Urb. Villafranca del Castillo, PO Box 50727, 28080 Madrid, Spain.\\
$^3$National Space Institute, Technical University of Denmark, Juliane Maries Vej 30, 2100 Copenhagen, Denmark.}

\date{Released 2002 Xxxxx XX}

\pagerange{\pageref{firstpage}--\pageref{lastpage}} \pubyear{2008}

\def\LaTeX{L\kern-.36em\raise.3ex\hbox{a}\kern-.15em
    T\kern-.1667em\lower.7ex\hbox{E}\kern-.125emX}

\begin{document}

\label{firstpage}

\maketitle

\begin{abstract}
 The timing properties of the 4.45~s pulsar in the Be X-ray binary system \gro~are examined using hard X-ray data from \integral~and \swift~during a Type~II outburst observed during 2008.  The orbital parameters of the system are measured and agree well with those found during the last known outburst of the system in 1995.  Correcting the effects of the doppler shifting of the period, due to the orbital motion of the pulsar, leads to the detection of an intrinsic spin-up that is well described by a simple model including $\dot{P}$ and $\ddot{P}$ terms of $-7.5\times 10^{-10}$~s~s$^{-1}$ and  $1\times 10^{-16}$~s~s$^{-2}$ respectively.  The model is then used to compare the time-resolved variation of the X-ray flux and intrinsic spin-up against the accretion torque model of \citet{1979ApJ...234..296G}; this finds that \gro~is likely located 12--22~kpc distant and that the surface magnetic field of the neutron star is $\sim 2\times 10^{12}$~G.  The shape of the pulse and the pulsed fraction shows different behaviour above and below 20~keV indicating that the observed pulsations are the convolution of many complex components.

\end{abstract}


\section{Introduction}
The number of known high mass X-ray binary systems (HMXB) has expanded in recent years with the advent of hard X-ray satellite telescopes with large fields of view enabling the serendipitous detection of these sources whilst in outburst.  In particular the number of HMXB systems comprising a neutron star (NS) with a Be star companion, which in most cases are only seen in brief outbursts following $\sim$~years of quiescence,  has increased to the point where the majority of HMXB known now are such systems (hereafter BeXRB).  Be stars are young and hot, with spectral Type B (or O) characterised by Balmer emission lines thought to be associated with circumstellar material shed due to the rapid rotation of the star, which may be near to its disruption speed \citep[see e.g.][]{2003PASP..115.1153P}.  The NS, in an elliptical orbit around the star, can accrete matter from an equatorial ``decretion disk'' that may be formed from the disrupted material, or from a strong stellar wind; this potentially leads to outbursts in X-rays seen at different orbital phases.  

The types of outburst can be separated into Type I and Type II \citep{1986ApJ...308..669S}:  Type I outbursts are seen to occur around the periastron of the binary orbit (although outbursts are not necessarily seen at every periastron) and are thought to be due to an increase in accretion onto the NS's surface when it is closest to the Be star where density of matter available is higher; Type II outbursts are much brighter than Type I, do not seem to be correlated with orbital phase and in many cases last for several orbital periods.   Pulsed X-rays are often seen from BeXRBs since they contain strongly magnetised NS; to date there are no known BeXRBs containing a confirmed black hole. Typical pulsar periods, $P$, for BeXRBs are in the range 1-100~secs and show a correlation with the binary periods, $P_{orb}$,  of 10-100 days \citep{1986MNRAS.220.1047C}.  During Type II bursts the pulsation rate of the pulsar is seen to increase ($P$ reduces, or frequency $\nu = 1/P$ increases) suggesting that significant angular momentum transfer from the star to the NS is happening; this phenomena is known as \emph{spin-up}.

The observation of Type II outbursts are particularly interesting for many reasons, especially those where a pulsar can be observed: with luminosities of 1-2~$\times 10^{38}$~ergs~s$^{-1}$ they are a factor of $\sim$~10 brighter than Type I outbursts and should allow good signal to noise even for systems located far away from Earth; the behaviour of the system over at least one full orbital period can be investigated, allowing the orbital charactersitics to be inferred from changes in  $P$; the matter accretion rate, $\dot{m}$, can be studied by relating the spin-up rate, $\dot{P}$, of the pulsar to the transfer of angular momentum from the Be star; a period of spin-down of the pulsar towards the end of a large outburst with large spin-up may be indicative that the pulsar has undergone a torque reversal caused by an interaction of the maximally spinning pulsar magnetosphere with the inner accretion disk - this would allow an estimate of the magnetic moment of the NS material to be examined; changes in the orbital phase averaged spectra may allow the Be star environment to be probed.

\subsection{\gro}
  The transient X-ray pulsar \gro~(also known as AX~J1749.1$-$2639) was discovered towards the Galactic Centre, by the \emph{CGRO}/BATSE satellite experiment \citep{scott}, following an outburst of 20--70~keV photons over the period MJD~49915--49978 (1995 July 17 -- September 18).  The BATSE detectors were large wide-angle uncollimated scintillator detectors, hence the outburst was detected only through a pulsed flux with period 4.45~s.  Any un-pulsed component was not measurable, in the absence of any imaging capability, and formed part of the background signal (along with all other sources in the FoV).  The pulsed flux was equivalent to $\sim$~30~mCrab at its peak.  The pulsation was observed to spin-up during the outburst at a rate  correlated with the pulsed intensity, reaching 38~pHz~s$^{-1}$ at the peak of the outburst.  A modulation of the period was found to correspond to a 29.8~day orbital period and led \citet{scott} to identify GRO~J1750-27 as a BeXRB undergoing a giant (Type II) outburst following the classification of \citet{1986MNRAS.220.1047C}.  \citet{scott} made an estimate of the distance to the source of at least 18~kpc from Earth.  Due to the non-imaging nature of the BATSE detectors this measurement is based on an assumed pulsed fraction of 30\% for \gro.

   \begin{figure}
   \centering
   \includegraphics[angle=270, width=9cm]{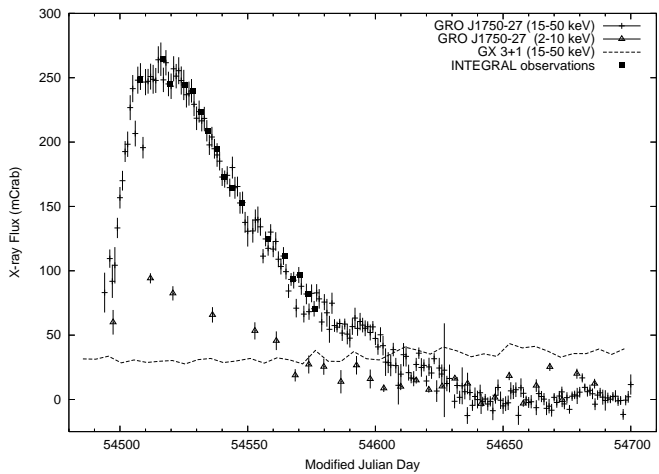}
   \caption{\gro~X-ray light curves from the \swiftbat~transient monitor webpage (http://swift.gsfc.nasa.gov/docs/swift/results/transients) and \emph{RXTE}/ASM (http://xte.mit.edu/ASM\_lc.html), shown by vertical crosses and triangles respectively.  Dark squares are overlaid on the BAT data to mark where simultaneous observations were made by \integral. The dashed line shows the 15--50~keV flux from the X-ray burster GX~3$+$1, which is located $18\arcmin$ from \gro~(average flux from 5~daily measurements is shown for clarity).}
              \label{fig:batlc}%
    \end{figure}

A second outburst was reported by the \swiftbat~transient monitor during an observation starting at 11:26:40 UT on 29th Jan 2008 (MJD 54494).  The source was first detected at $0.031 \pm 0.005$~ cts/cm$^2$/sec (140~mCrab) in the 15-50 keV energy range of the BAT detector \citep{batatel}.  The outburst is shown in Fig.~\ref{fig:batlc} and consists of a sharp rise to the peak flux in $\sim$~30~days and a slower decay over the following $\sim$~120~days.  The XRT soft X-ray telescope, also onboard \swift, has given the currently most accurate position for \gro~ as $\rmn{RA}(2000)=17^{\rmn{h}} 49^{\rmn{m}} 13\fs1$,$\rmn{Dec.}~(2000)=-26\degr 38\arcmin 37\farcs3$ with an uncertainty of 5\arcsec \citep{2008GCN..7274....1B}\footnote{A position of \gro, measured with the \emph{ROSAT} X-ray telescope, is given in \citet{scott} as $\rmn{RA}=17^{\rmn{h}} 49^{\rmn{m}} 12\fs7$, $\rmn{Dec.}~=-26\degr 38\arcmin 36\farcs~(J2000)$ with 4\arcsec~accuracy, and is cited as \emph{Dennerl and Trumper, A\&A, 1997, (in press)}; this paper, or any later one with the same authors, was not found in a literature search.}.  Note that the location of \gro~is $\sim$~18\arcmin~from the bright X-ray burster GX~3$+$1, which requires that observations must be made with instruments with good angular resolution.  No optical counterpart brighter than $R \simeq 17$~mag has been reported, although the assumed distance to \gro~of 18~kpc and its location of ($l$,$b$) = (2\fdg4, +0\fdg5) place it some distance beyond the Galactic Centre, where optical observations are very difficult due to absorption.

\begin{figure}
   \centering
   \includegraphics[width=9cm]{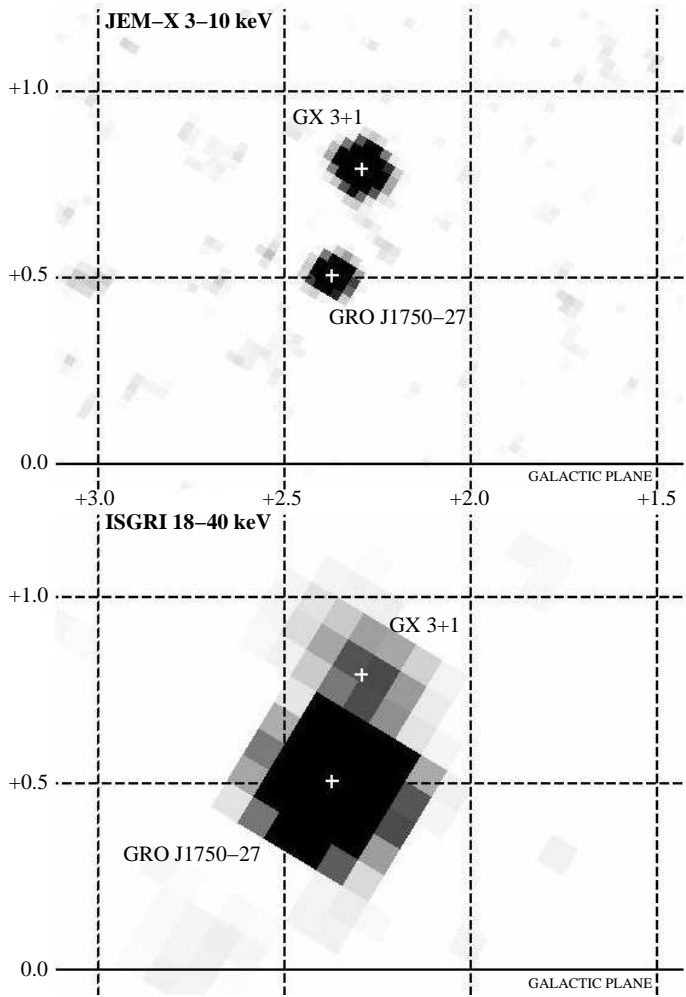}
   \caption{Images of the \gro~region taken with \integral~during revolution 654 (2008 Feb 20; MJD~54516) in the 3--10~keV range with JEM-X1 (top) and 18--40~keV with ISGRI (bottom).  The grid shows Galactic coordinates while the grey-scale represents  statistical significance.  White crosses mark the respective source positions.  In the JEM-X image \gro~ is detected with a peak significance of $36\sigma$ (corresponding flux 133~mCrab);  the peak of GX~3$+$1 is $60\sigma$ (235~mCrab).  In the ISGRI image \gro~ is detected with a peak significance of $117\sigma$ (194~mCrab);  the peak of GX~3$+$1 is $13\sigma$ (21~mCrab)}
              \label{fig:jximg}%
    \end{figure}

This outburst was also observed by the ESA X-ray/$\gamma$-ray satellite \integral~during Galactic Bulge monitoring observations \citep{intatel1, erik}.  {\correction The monitoring programme consisted of $\sim$~12~ksec observations of the Galactic Bulge repeated approximately once per orbital revolution of \integral~(the orbital period of \integral~is $\sim 72$~hours) and continued throughout the brightest part of the outburst. } Figure~\ref{fig:jximg} {\correction shows images taken with the \integral~instruments during the peak of the outburst.  The JEM-X instrument has finer angular resolution than ISGRI and hence \gro~and  GX~3$+$1 are clearly resolved in the low energy image.  Nevertheless the two sources are also resolved with ISGRI and, due to the relatively softer spectrum of GX~3$+$1, \gro~is many times brighter at higher energies}.

{\correction The all sky monitor (ASM) data from the \emph{RXTE} satellite 
were also inspected to give further coverage of the outburst in the 2--10~keV range (see Fig.~\ref{fig:batlc}).  Individual ASM observations of \gro~have low signal to noise ratios, since the low energy flux was approximately a third of that at higher enrgies and because the measurements were based on typically only a few hundred seconds of exposure;  the observations were also sparcely spaced, with several days between each.  This meant that it was necessary to rebin the ASM light curve in order to reduce the uncertainty on the flux measurements to an acceptable level.  Finally, the ASM data are affected by contamination from nearby sources, therefore this data was not considered further}.

Following the 2008 outburst, the data from \integral~and \swift~were obtained, allowing this fascinating source to be revisited with the advantage of hard X-ray data from simultaneous observations with imaging capability and the benefit of long quasi-simultaneous exposures.

\section{Data}
\subsection{\integral}
The \integral~satellite is a X-ray/$\gamma$-ray observatory that was launched by ESA on 17 October 2002 \citep{2003A&A...411L...1W}.  It was designed with several instruments to offer fine spectroscopy and imaging in the 3~keV--10~MeV range.  Two instruments in particular are of interest for this work: above 18~keV, IBIS/ISGRI \citep{2003A&A...411L.131U, 2003A&A...411L.141L} provides hard X-ray imaging in a large (30\degr) field of view with angular resolution $\sim$~12$^{\prime}$, point source location accuracy $\sim 3^{\prime}$, sensitivity $\sim$~18~mCrab for a 5$\sigma$ source seen in an 1.8~ksec observation in the 20--60~keV range \citep[e.g.][]{erik}; the JEM-X soft X-ray monitor \citep{2003A&A...411L.231L} consists of two identical coded-mask gas detectors sensitive to X-ray photons in the range  3 -- 25~keV within a 5\degr~field of view and an angular resolution of 30\arcsec.  Only the JEM-X1 detector was operational during the observations described here.

\integral~makes observations of the Galactic Bulge, approximately every 3 days during the monitoring of the Galactic Bulge \citep{erik} and is ideal for following up $\sim$~month long bursts.  The coded mask instruments JEM-X and ISGRI have been used to measure the pulsations from \gro~in the range 3--70~keV.  The data were reduced with the standard Offline Standard Analysis software version 7 (\osa)\footnote{\osa~is available from the ISDC at http://isdc.unige.ch} provided by the Integral Science Data Centre \citep[ISDC;][]{2003A&A...411L..53C}.

Both ISGRI and JEM-X have good timing charactersistics, depending on the brightness of the source in question.  Investigation of the 4.45~s pulse from \gro~is easily within the capabilities of both \integral~ instruments. 

\subsection{\swift}
The \swiftbat~instrument on the \swift~satellite \citep{2004ApJ...611.1005G,2005SSRv..120..143B} is a wide angle coded mask system in a Low Earth Orbit, designed primarily to detect Gamma Ray Bursts (GRB).  As such it has a very large field of view, approximately 2~sr, in order to maximise its chance of detecting a randomly occurring GRB.  This means that during the course of its normal operations about 15\% of the sky is observed and so most areas of the sky have some exposure during each 90~minute orbit.  In survey mode, images are produced in histogrammes of approximately 5~minutes length, allowing a light curve of the outburst to be produced with excellent time coverage as seen in Fig.~\ref{fig:batlc}.  Of course, the survey image histogrammes are of no use for timing investigations on time-scales less than the histogramme duration.  However, 1~second count rate data from the whole detector plane is available, which allows strong periodic signals of more than a few seconds length to be observed.  In this respect, the observations from BAT are comparable to those made with BATSE in 1995; strong pulsed signals should be detectable, although no measurement of absolute fluxes will be possible from the rate data.

\section{\integral~spectrum at the peak of the outburst}
\label{sec:spec}
{\correction Exisiting X-ray instruments are unable to use reflection or refraction in optical materials to focus photons with energies more than a few keV (note, however, that several new missions are planned that aim to achieve this in the future with various methods: Simbol-X, NuStar, NeXT, GRI)}.  The instruments onboard \integral~rely on coded masks to cast a shadow on a position sensitive detector, which can then be used to reconstruct the direction of the incident light.  Some contamination of the flux measurement for a particular object is  therefore unavoidable in observations of fields containing many sources.  Spectral analyses of objects in the Galactic Centre will therefore be challenging and especially so for \gro~given its proximity to, and the relative brightness of, GX~3$+$1 (see Fig.~\ref{fig:jximg}).

In particular, the \osa~tools for spectral analysis with JEM-X are optimised for a single on-axis source and for fields containing many sources it is recommended to make spectra from fluxes taken from images made in discrete energy ranges\footnote{See \emph{Known Issues} in the JEM-X analysis cookbook at http://isdc.unige.ch/?Soft+download}.  This method is thought to be more immune to contamination of the spectra than the standard method, although it is not practical to make images in bands less then $\sim$~1--2~keV wide.  A comparison of spectra made by both methods, for the peak of the outburst during revolution 654, showed that they were consistent above 5~keV; at lower energies there was a clear overestimate in the flux using the standard \osa~spectral extraction method.

For higher energy spectral analysis with ISGRI the detrimental effects of other sources are somewhat reduced since, in revolution 654, \gro~is brighter than GX~3$+$1 above 18~keV by about a factor of ten and is also the brightest source in the whole field of view by a factor of two.  

The broadband spectrum during the peak of the outburst was therefore calculated in the 5--70~keV range and fit with a cut-off power law:

\begin{equation}
A(E) = K \left(\frac{E}{\rmn{1 keV}}\right)^{-\Gamma} e^{(-E/E_{c})}
\end{equation}

\begin{figure}
  \centering
  \includegraphics[angle=270,width=9cm]{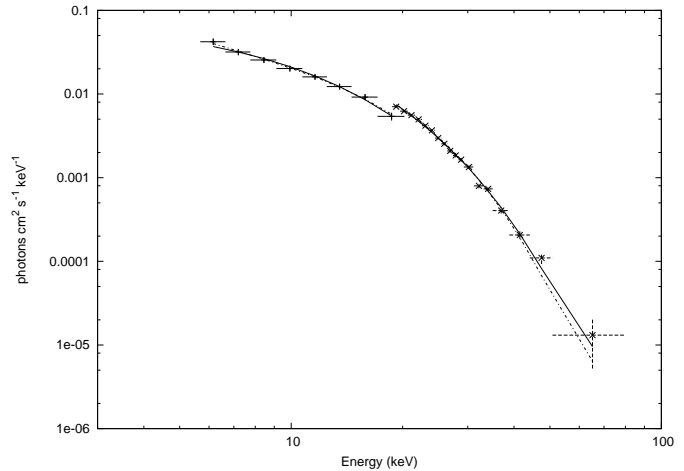}
  \caption{The \integral~spectrum of \gro~during the peak of the outburst in revolution 654.  Measurements with JEM-X1 cover the 3--20~keV range (vertical crosses), whilst ISGRI data is taken from 18--50~keV (diagonal crosses).  The lines show models which have been simultaneously fitted to both data sets as described in section~\ref{sec:spec}: comptonisation (continuous line); cut-off power-law (dashed-dotted line).}
  \label{fig:spec}%
\end{figure}

\begin{figure}
   \centering
   \includegraphics[width=9cm]{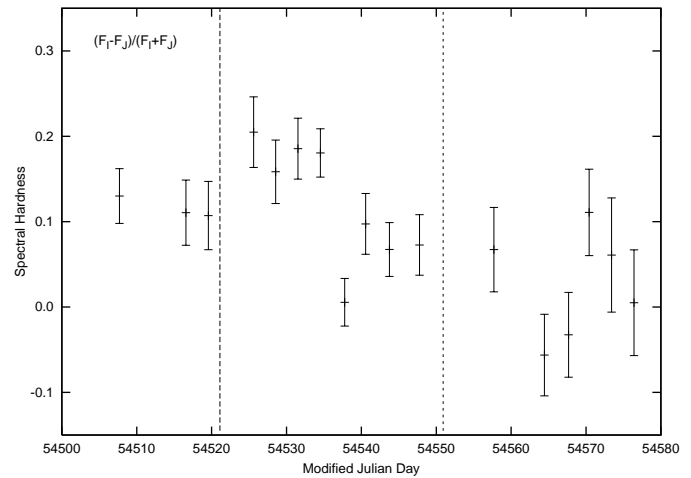}
   \caption{\integral~ measurements of the hardness ratio, $\frac{(F_{I}-F_{J})}{(F_{I}+F_{J})}$, where $F_{I}$ is the 20--50~keV ISGRI count rate and  and $F_{J}$ is the 5--20~keV JEM-X count rate.  The dashed vertical lines show the times of periastron of the \gro~system.}
              \label{fig:hr}%
    \end{figure}

{\correction In order to investigate the energy spectrum of \gro, the data from JEM-X1 and ISGRI were fitted simultaneously with a phenomenological model, namely a cut-off power-law function.  The spectrum can be seen in Fig.~\ref{fig:spec}.  Systematics of 3\% were added to JEM-X and 2\% added to ISGRI and an inter-instrument calibration factor, defined as a factor applied to the ISGRI spectral response with JEM-X held constant,  was also included as a free parameter.  This gave a chi-squared value of $\chi^{2}/dof =  28.6/22$ and returned values of $\Gamma = -0.15\pm0.3$ and $E_{c} = 6.0^{+0.5}_{-0.4}$~keV respectively (90\% confidence) and an instrument calibration factor of 1.49. Assuming that most of the energy output of the source is in the range 0.1--100~keV gives a total flux of 6.5$\times 10^{-9}$~ergs~cm$^{-2}$~s$^{-1}$, although the uncertainty on this is $\pm$30\% due to the variation on the measurement of $\Gamma$ alone.  

A physical model, ascribing the spectral behaviour of the source to the inverse Compton scattering of soft photons by a thermal plasma of electrons, was also fitted to the data.  The {\sevensize COMPTT}  model of \citet{1994ApJ...434..570T} returned a reduced chi-squared value of 1.24 and the following values: the plasma temperature  $4.6\pm0.1$~keV; the plasma optical depth $6.4^{+0.9}_{-0.7}$; the the soft photon temperature was fixed at 0.1 keV as it lies far outside of the range covered by our spectral data.  The inter-instrument calibration factor was 1.40.   This model follows essentially the same form as the cut-off power-law, as can be seen in Fig.~\ref{fig:spec}, and suggests that the X-ray spectrum is due to the scattering of soft X-ray photons in an optically thick plasma.

In both cases, the calibration factor returned was not equal to unity. This shows that there are some systematic effects in the spectrum that are not fully understood when dealing with moderately bright sources in crowded fields.  Since the fields of view of ISGRI and JEM-X are not the same size, the backgrounds of both instruments will be dependent on different objects.  The objects contained in both fields of view will also not have the same effects on each instrument, due to the large range of the X-ray spectra of the objects in the Galactic Bulge;  each source may have a different relative brightness to \gro~in the energy bands of the two instruments, and will therefore be of changing importance to both background estimations.

The detailed analysis of the \gro~spectrum is clearly complicated even when the source is at its brightest; understanding the instruments' spectral response during the observations, and hence the spectral behaviour of \gro~ throughout the outburst and its variation with pulsar phase, will be the subject of future work.}  In order to investigate the spectral behaviour, without model dependent spectral fits, simple spectral hardness ratios have been calculated from the total count rate fluxes measured by \integral~during the outburst: 

\begin{equation}
h =  \frac{(F_{I}-F_{J})}{(F_{I}+F_{J})}
\end{equation}
where $F_{I}$ is the 20--50~keV ISGRI count rate and  and $F_{J}$ is the 5--20~keV JEM-X count rate.  Figure~\ref{fig:hr} shows the time evolution of the broad band spectral hardness in the 5--20 and 20--50~keV ranges for the total (pulsed+unpulsed) data.  There is a gradual softening of the total flux from \gro~during the outburst, with some indications of more complex variability.

\section{Hard X-ray Timing Analysis}

\subsection{ISGRI pulsation detection at 18--40 keV}
\label{sec:timingisg}
The \osa~tools \emph{ii\_light} and \emph{barycent} were used to create 18--40~keV light curves, with 0.5~s binning, and to correct the time stamps of the results for the orbital motion of the \integral~satellite, for each set of Galactic Bulge monitoring observations.

The ISGRI light curves analysed here are made from several fixed pointings separated by slews and, as is common in high-energy astrophysics observational analysis, may contain irregularly distributed time data affected by telemetry gaps etc.  Therefore the Lomb-Scargle method  \citep{1976Ap&SS..39..447L, 1982ApJ...263..835S}, which is robust against these effects, is used to search for periodic signals.  The error on the angular frequency of any detected pulsation is given by the formulations of \citet{1986ApJ...302..757H}:

\begin{equation}
\delta \omega = \frac{3\pi\sigma_{N}}{2\surd N T A}
\label{eqn:LSper}
\end{equation}

where the amplitude $A$ is  calculated thus,

\begin{equation}
A=2\sqrt{\frac{z_{0}\sigma_{s}^{2}}{N}}
\label{eqn:LSerr}
\end{equation}

$\sigma_{N}$ is the standard deviation of the noise in the lightcurve, estimated to be the same as the standard deviation of the lightcurve, $\sigma_{s}$, for a weak pulsed signal; $z_{0}$ is the power of the pulsation from the periodogramme; $N$ is the number of data points; $T$ is the total length of the data set.

In each of the 17 Galactic Bulge monitoring observations the Lomb-Scargle periodograms showed a significant peak at a frequency of $\nu$$\sim$0.22 Hz.  This corresponds to a period of $\sim$4.45~s which is in agreement with the observed spin period identified by \citet{scott}.  Table~\ref{tab:isgri} lists the detected pulse period and the associated error for each observation. The ISGRI data clearly show periodic signals which can be measured with very high accuracies, typically $\sim$~few parts in $10^{5}$.

\subsection{ISGRI measurements of pulsar spin-up}
\label{sec:spinisg}
\begin{figure}
   \centering
   \includegraphics[width=9cm]{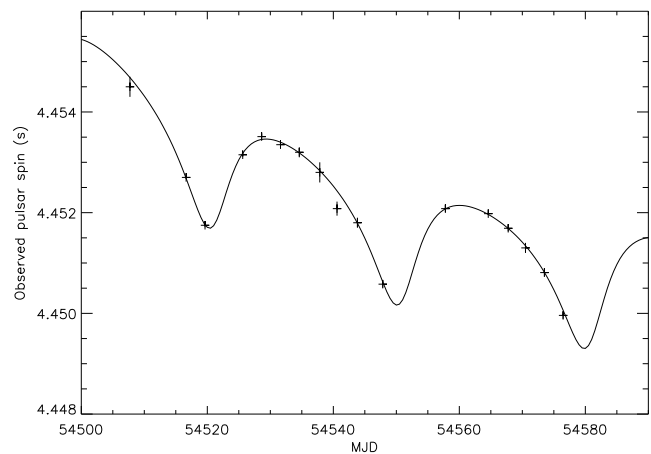}
   \caption{\gro~ orbital solution based on ISGRI measurements (crosses). Note in most cases the error bars on the period are much smaller than the size of the crosses.}
              \label{fig:orbitsol}%
    \end{figure}

The \gro~pulse period, $P$, as a function of observation time is shown in Fig.~\ref{fig:orbitsol} and clearly indicates that the observed spin period evolves with time.  While the oscillatory motion may be explained by a doppler shifting of the spin-period by the binary orbit of the system there is also an additional component resulting in an intrinsic spin-up of the pulsar.

The observed data was fit with a simple spin-up model for the pulsar given by:
\begin{eqnarray}
	P(t) & = & P(t_0) + \dot{P}(t-t_0) - \frac{\ddot{P}(t-t_0)^2}{2}
\label{eqn:spinup}
\end{eqnarray} 
This model is convolved with a standard orbital model assuming the orbital parameters measured by \citet{scott}, excluding the orbital period.  This combined model when fit to the data resulted in a chi-square value of $\chi$$^{2}$/dof = 12.78/13.  An orbital period of 29.806~$\pm$~0.001 days was found which is compatible with that previously observed; there is also a clear measurement of $\dot{P}$ and $\ddot{P}$.  The details of the fit parameters and the orbital parameters of \citet{scott} are given in Table~\ref{tab:orbit}.  The resultant best fit model is presented by the solid line in Figure~\ref{fig:orbitsol}.

Refitting the data without assuming any of the parameters of  \citet{scott}  yields an acceptable fit.  However, whilst the orbital parameters are compatible with the previous observation they have larger associated errors; therefore the prefered fit is based on the original parameters.  Removing the $\ddot{P}$ parameter results in an unacceptable fit, $\chi$$^{2}$/dof = 50.04/14.  An ftest on these fit results shows that the addition of $\ddot{P}$ is highly significant, with a chance improvement probablility of only $3\times10^{-5}$.  Removing both the $\dot{P}$ and $\ddot{P}$ terms returns an even poorer fit by many orders of magnitude: $\chi$$^{2}$/dof~$\sim$~6000/15.  Hence the intrinsic spin-up model of equation~\ref{eqn:spinup} is determined with a very high level of confidence.

   \begin{table}
     \caption[]{18--40~keV pulsations detected from \integral/ISGRI observations of \gro.}
     \label{tab:isgri}
     \begin{tabular}{c c c l l}        
       \hline
       Revolution & MJD & Exposure (ks)& Period (s) & $\Delta$Period (s)\\
       \noalign{\smallskip}
       \hline
       \noalign{\smallskip}
0651  &        54507.687 &      5 &  4.4545   &    0.0002 \\ 
0654  &        54516.572 &     12 &  4.45270  &    0.00003 \\ 
0655  &        54519.562 &     12 &  4.45175  &    0.00004 \\ 
0657  &        54525.545 &      7 &  4.45315  &    0.00004 \\ 
0658  &        54528.538 &     12 &  4.45351  &    0.00006 \\ 
0659  &        54531.530 &     13 &  4.45335  &    0.00007 \\
0660  &        54534.522 &     12 &  4.4532   &    0.0001 \\ 
0661  &        54537.773 &     12 &  4.4528   &    0.0002 \\ 
0662  &        54540.509 &     10 &  4.45208  &    0.00014 \\ 
0663  &        54543.752 &     12 &  4.45180  &    0.00010 \\ 
0664  &        54547.759 &     14 &  4.45058  &    0.00007 \\ 
0667  &        54557.708 &     12 &  4.45208  &    0.00007 \\ 
0670  &        54564.432 &     24 &  4.45198  &    0.00002 \\ 
0671  &        54567.682 &     12 &  4.45169  &    0.00006 \\ 
0672  &        54570.413 &     10 &  4.45130  &    0.00010 \\ 
0673  &        54573.403 &     11 &  4.45081  &    0.00008 \\ 
0674  &        54576.395 &     12 &  4.44996  &    0.00007 \\ 
       \noalign{\smallskip}
       \hline
     \end{tabular}
   \end{table}

\begin{table}
\centering
\caption{The orbital parameters for GRO J1750$-$27 derived from \citet{scott} and from the analysis of ISGRI 20--40 keV data.}
\begin{tabular}{|l|cl}
\hline
Scott et al. (1997) &  & \\
\hline
\hline
Orbital period	& P$_{orb}$ & 29.817$\pm$0.009 days\\
Projected semimajor axis	&  a$_{x} \sin{i}$ & 101.8$\pm$0.5 lt-s \\
Eccentricity		& \textit{e}	&	0.360$\pm$0.002\\
Longitude of periastron	&	\textit{$\omega$}	& 206$^{\circ}$.3$\pm$0$^{\circ}$.3\\
Orbital epoch	&	T$_{periastron}$	& MJD 49931.02$\pm$0.01\\
\hline
& &\\
\hline
This work & & \\
\hline
\hline
Orbital period	& P$_{orb}$ & 29.806$\pm$0.001 days \\
Spin period (@ MJD 54516.635)	& $P$ & 4.45349$\pm$0.00002 days \\
1$^{st}$ derivative of $P$	& $\dot{P}$ & $-$7.5$\pm$0.3 $\times$ 10$^{-10}$ s~s$^{-1}$ \\
2$^{nd}$ derivative of $P$	& $\ddot{P}$ & 1.0$\pm$0.1 $\times$ 10$^{-16}$ s~s$^{-2}$ \\
\hline
\end{tabular}
\label{tab:orbit}
\end{table}

\subsection{Pulsation measurements with JEM-X}
Following similar methods to the ISGRI analysis, data from the soft X-ray monitor onboard \integral, JEM-X1,  were reduced with \osa~for a broad 5--20~keV energy band. The analysis of the JEM-X data returned values of $P$ consistent with the ISGRI results for all contemporaneous observations, albeit at lower statistical significance.  In several of the earlier observations significant power was also seen at $\sim$~2.22~s, corresponding to half of the ISGRI period.  Inspection of the phase-folded JEM-X light curves showed that below 20~keV the pulse amplitude was weaker and in some cases showed a double peaked profile. This is investigated in section~\ref{sec:profile}

\subsection{Searching the \swiftbat~data for a periodic signal}

\label{sec:timingbat}

The $\sim$~2~sr field of view of BAT means that many observations are made of the source location; if suitable power spectra can be produced then the spin history will be well covered in time.  Since the survey mode data, from which images can be produced, are integrated over $\sim$~5~mins they are not usable for timing studies of \gro.  The only BAT data with suitable time resolution is the detector plane count-rate, in which the total flux on the whole detector plane is read every second.  Since no positional information is available from these data, any pulsed signal from \gro~is contaminated by the signals from all other sources in the very large field of view.  This can be effectively considered as an increase in the background noise which will appear non-statistical.  No time dependent background correction is available for the count-rate data in the current Swift standard analysis and consequently it has not been possible to extract reliable pulsed fluxes.  The BAT does, however, have the advantage of many observations of GRO J1750$-$27 throughout the outburst.

Data were obtained for all pointings between MJD 54480 and 54640 (2008 January 15 and June 23) where the position of \gro~was within 30\degr~of the pointing direction.  This was to obtain a sub-set of data not too badly affected by noise towards the edge of the detector plane and so that the source was detected with a reasonable sensitivity. The data were broken up into 50~ks segments, which were defined consecutively from the start of the observation, and were each searched for a periodic signal.  Due to the low Earth orbit of Swift and its pointing strategy the data are generally arranged in a few short sections of $\sim$~1000~seconds duration within the time bin. 

To improve the chances of detecting the pulsar spin period a Bayesian statistical approach is applied to the data set.  The generalised Lomb-Scargle periodogram method of \citet{2001AIPC..568..241B} is used with the period range of interest restricted to the narrow range 4.445--4.460~s i.e. tightly bounding the periodicity detected by ISGRI.

\begin{figure*}
  \centering
 \includegraphics[angle=90, width=18cm]{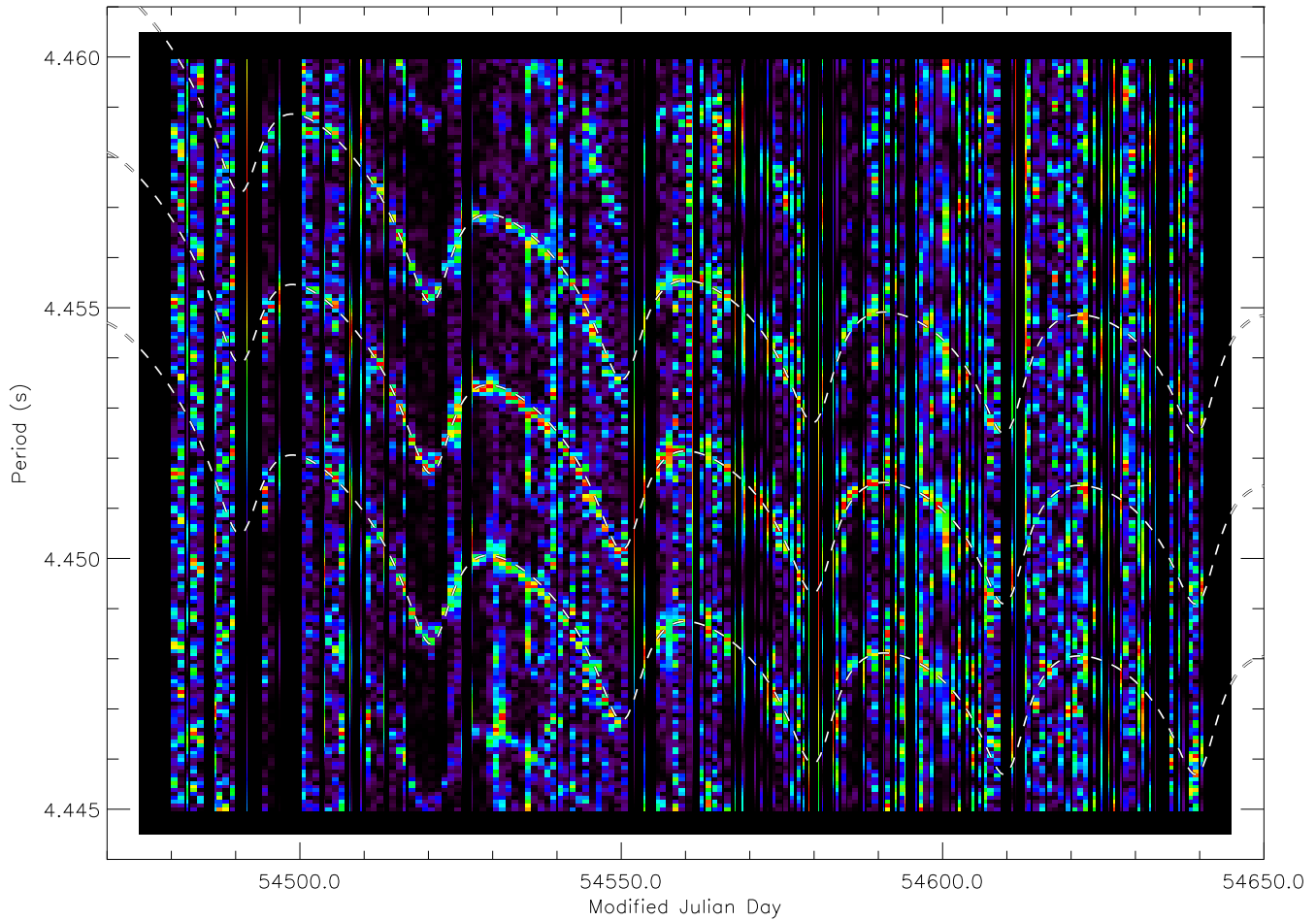}
\caption{Dynamic periodogrammes measured from 50~ksec sections of \swiftbat~detector plane count-rate data for the period MJD~54480--54640 (2008 January 15 -- June 23) using the Bayesian analysis method outlined in section~\ref{sec:timingbat}.  Pulsation probability is represented with a rainbow colour scale with red representing high confidence of a pulsed signal.  The central dashed line shows the orbital solution for the pulse variation measured from the ISGRI data (see section~\ref{sec:timingisg}); the upper and lower dashed lines show the model offset by $\pm$~3.4~ms. The ISGRI orbital model is in very good agreement with the BAT observations, no scaling or normalising of the ISGRI model has been performed. The offset lines agree with the secondary tracks caused by the aliasing of the pulsed signal with the \swift~spacecraft orbital period.}
\label{fig:batspec}%
\end{figure*}

Figure~\ref{fig:batspec} shows the resulting periodograms displayed as a function of time and coloured for the probability density at each period.  The highest probability density corresponds to periods which are consistent with the ISGRI measured pulses and the fitted period model; this is shown by the central dashed line in Figure~\ref{fig:batspec}.  Also evident are side lobes of lower probability which follow the same time evolution; these are offset from the main track by integer units of a constant, k = $\pm$3.4~ms.  If this effect is caused by the beating of the pulsation with another periodic signal is given by the following:
 
\begin{equation}
\frac{1}{P_{beat}}  =  \frac{1}{P} - \frac{1}{P + k}
\end{equation}

\noindent Inserting $P = 4.450$~s gives P$_{beat} \sim$~97 minutes, which corresponds approximately to the orbital period of the \textit{Swift} satellite.  The secondary tracks are therefore very likely to be due to aliasing with subtle effects on the count-rate or spacecraft orientation occurring on the \swift~geo-orbital period.

Figure~\ref{fig:batspec} shows that ISGRI derived orbital solution is valid throughout the outburst and can be reliably extrapolated to times after the ISGRI observations ended at MJD~54576.  The first BAT survey mode data available for the outburst, as shown in Fig.~\ref{fig:batlc}, is on MJD~54494; detector plane rate data is available before this time, although there does not seem to be any strong evidence for a pulsed signal before $\sim$~MJD~54490.  

As $\dot{P}$ is a tracer of the mass accretion rate, $\dot{m}$, $\ddot{P}$ can be used as an indicator of how $\dot{m}$ changes
i.e. the rate at which the mass contained within the accretion disc is being
exhausted.  Consequently, we can define a simple timescale for the survival
of the accretion disc, $\tau$, as:

\begin{eqnarray}
\tau & = & -\frac{\dot{P}}{\ddot{P}} 
\end{eqnarray}

\noindent This gives $\tau =$ 87~$\pm 9$~days based upon the ISGRI
measurements shown in Table~\ref{tab:orbit}.  Extrapolating from the time of peak of
the outburst this implies that the outburst should end at $\sim$MJD 54611.

\subsection{Pulse Profile}
\label{sec:profile}
\begin{figure*}
  \centering
  \includegraphics[width=15cm]{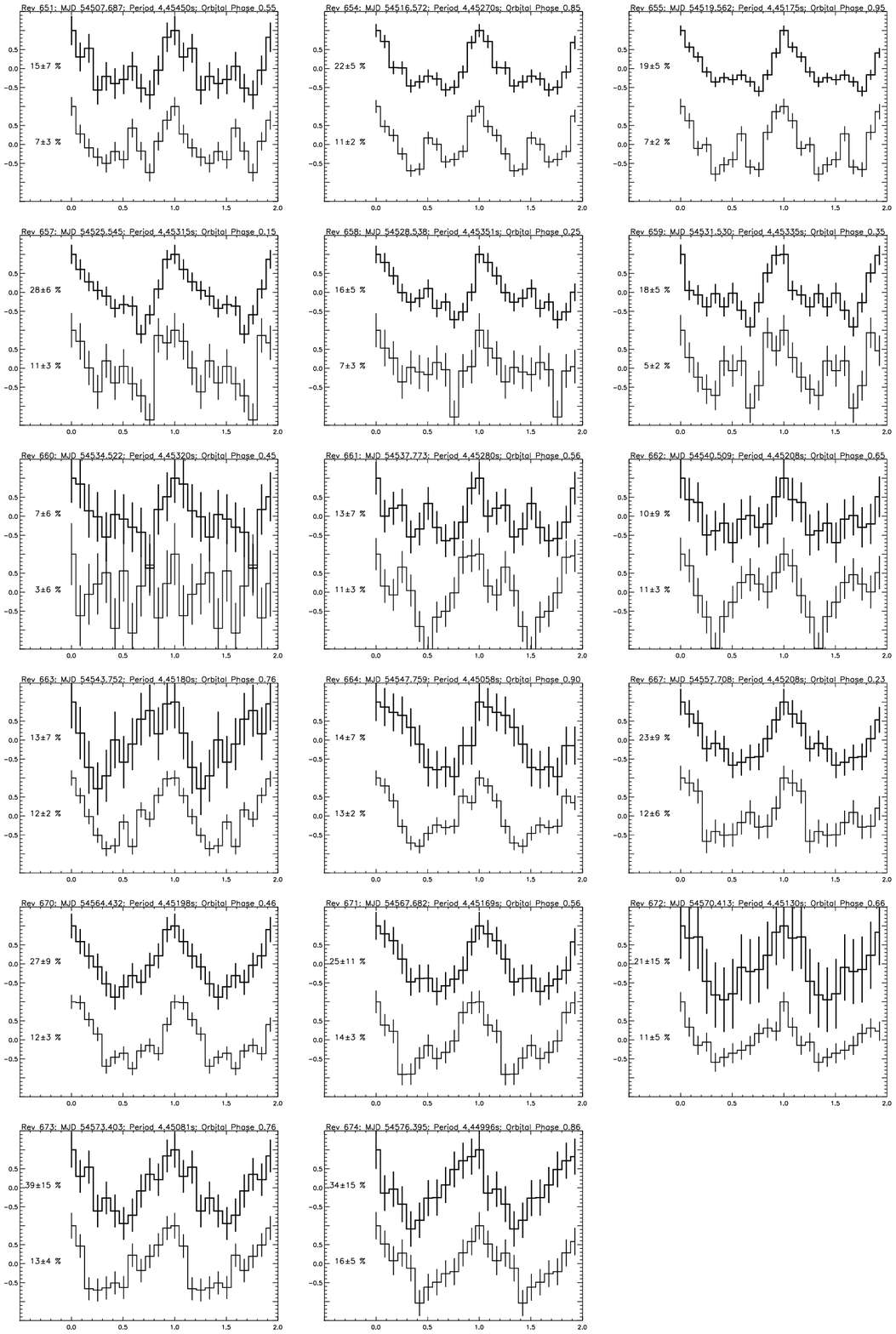}
  \caption{Relative pulse profile diagrammes for different \integral~observations of the 2008 outburst of \gro: ISGRI 20--50~keV (top, thickest); JEM-X 5--20~keV (bottom, thinnest). The data were folded on the periods listed in Table~\ref{tab:isgri} and are indicated at the top of each plot along with the start date of the observation and the estimated orbital phase ($0 =$ periastron) at that date.   For clarity, the profiles are repeated for two phases.  The numbers to the left of the profiles indicate the pulsed fraction and uncertainty.}
  \label{fig:profiles}%
\end{figure*}

The shape of the pulse profiles has been measured over the 5--50~keV range with \integral.  As seen in section~\ref{sec:timingisg} the ISGRI timing results are highly accurate with strong power spectra leading to uncertainties on $P$ of typically 200~$\mu$s.  Phase diagrammes for the pulsar were made from the ISGRI 20--50~keV data from each revolution by folding on the measured period.  An offset was then added to the phase diagram so that the brightest phase bin was centred at phase $=0$.  The period and epoch defined by ISGRI were then used to fold the JEM-X data in 5--25~keV band in order to make comparable phase diagrammes, which are shown in Figure~\ref{fig:profiles}.

In general the shapes of the pulses seem to be more complex in shape below 20~keV than they are above 20~keV;  below 20~keV, there are clear signs of a double peaked pulse profile.  A secondary peak, approximately 50\% the size of the primary peak, is especially noticeable in revolutions 654 and 655.  In the same revolutions the pulse profile also shows signs of a secondary peak above 20~keV, although its relative size is much reduced.  The pulse appears weak in revolutions 660--662; in revolution 660 the pulse can barely be detected at all.  It is interesting to note that these three observations straddle the apastron of the system, when the separation between Be star and NS is largest and presumably the rate of accretion from any circumstellar material would be at a minimum.  The suggestion is therefore that there is a connection between the orbital phase and the  strength or shape of the pulsation.  However, the observations of revolutions 670--672 are also taken at approximately the same orbital phases, and here a very clear sinusoidal pulse is visible.  Unfortunately, determining any connection with the behaviour of \gro~and the orbital phase would certainly require observing an outburst over a larger number of orbital cycles and is not possible here.  The variation in the pulse shapes from observation to observation show that it would not be appropriate to attempt to improve the signal to noise by making an average of many pulses.

\subsection{Pulsed Fraction}
\label{sec:pf}
\begin{figure}
   \centering
   \includegraphics[width=9cm]{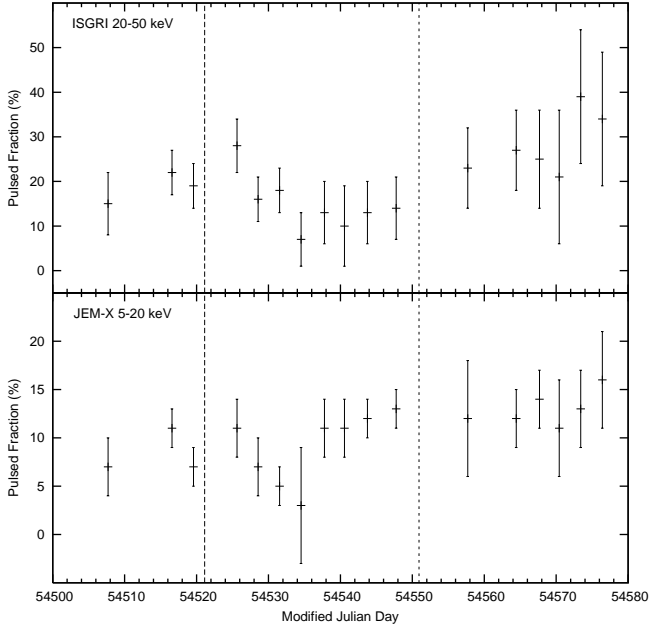}
   \caption{\integral~ measurements of the pulsed fraction during the outburst. The dashed vertical lines show the times of periastron of the \gro~system.}
              \label{fig:pulsedfrac}%
    \end{figure}

The evolution of the pulsed fraction as defined in \citet{1997ApJS..113..367B}, namely the ratio of the integrated peak-trough flux divided by the mean (pulsed+unpulsed) flux, is shown in Fig.~\ref{fig:pulsedfrac}.   The pulse profiles in Fig.~\ref{fig:profiles} were rebinned by a factor of two in order to calculate these values.  The pulsed fraction in the 5--20~keV band varies between 5--15~\%; above 20~keV the pulsed fraction is higher and has a larger variation, $\sim$~10--40~\%.  Both the 5--20~keV and 20--50~keV data show similar evolutions of the pulsed fraction with a slow rise in the first few observations, followed by a minimum at revolution 660.  In the 5--20~keV band the pulsed fraction seems to recover quickly to around 12~\% in revolution 661 and remains approximately constant from then on; this behaviour is quite different above 20~keV, where the pulsed fraction increases slowly from revolution 661 towards the end of the \integral~observations.

\section{Discussion}
The pulsar spin-up and orbital parameters of the BeXRB system \gro~have been measured to a high precision from hard X-ray observations with \integral~and \swift~(see Table~\ref{tab:orbit}).  These were found to be in good agreement with the measurements made by \citet{scott} during the previous known outburst from this system in 1995.  Assuming a cut-off power law model to describe the spectrum measured with \integral~during the peak of the outburst and extrapolating to 0.1--100~keV gives a total flux from the source of 6.5$\times 10^{-9}$ergs~cm$^{-2}$~s$^{-1}$; assuming that during the peak of the outburst the source is radiating at the Eddington luminosity, 1.8$\times 10^{38}$ergs~s$^{-1}$, puts an estimate on the distance to the source of around 15~kpc although this value is very uncertain due to the difficulties measuring and then extrapolating the 5--50~keV spectrum and also assumes spherically symetric accretion onto the NS.

The observed rapid spin-up of X-ray pulsars in outburst is proportional to the rate of angular momentum transfer from the donor star, which is in turn proportional to the rate of matter transfered to the NS through accretion, and also proportional to the X-ray luminosity of the source \citep[e.g][]{1972A&A....21....1P}.   In binary systems with an accretion disk, the rate of spin-up is curtailed by the threading of the accretion disk with magnetic field lines from the NS, which causes a torque to be imparted to the NS and acts to reduce the rate of spin-up.  This model was first proposed by \citet{1979ApJ...234..296G} and although other more detailed models have been proposed \citep[e.g][]{1987A&A...183..257W, 1995ApJ...449L.153W} they are largely consistent with the original.  \citet{1979ApJ...234..296G} suggest that for a NS with radius $1\times 10^{6}$~cm and given magnetic moment, $\mu$, the value of $\dot{P}$ is dependent only on the quantity $PL^{3/7}$, where $L$ is the bolometric luminosity of the source.  The changing luminosity can be estimated from the \swiftbat~light curve (fig \ref{fig:batlc}), which has excellent time coverage of the outburst and consists of average flux measurements taken daily.

The BAT count rate was converted into erg~cm$^{-2}$s$^{-1}$ by comparing with the many \swiftbat~observations of the Crab Nebula; the average Crab Nebula count rate in 15--50~keV is 0.22~counts~cm$^{-2}$s$^{-1}$.  In the same energy range, the Crab nebula was estimated to be  $1.6\times 10^{-8}$~erg~cm$^{-2}$s$^{-1}$, assuming a photon spectrum of $dN/dE = 10E^{-2.05}$ \citep{2007hsaa.book.....Z}. In order to find a bolometric luminosity, it is assumed that most of the energy output from \gro~is in the range 0.1--100~keV.  The spectrum described in section~\ref{sec:spec} returns a 15--50~keV flux that is approximately a factor of 3 less than the total flux.  Hence the BAT count rates can be converted to total energy output in ergs by scaling by the factor $3\times(1.6\times 10^{-8}/0.22)$.  This leads to a simple expression for the luminostiy in units of $10^{37}$~ergs~s$^{-1}$, $L \approx 2.6 F_{\rmn{BAT}} d^2$, where  $F_{\rmn{BAT}}$ is the BAT measured count rate in the 15--50~keV band and $d$ is the distance to the source in kpc. 

\begin{figure}
   \centering
   \includegraphics[width=9cm]{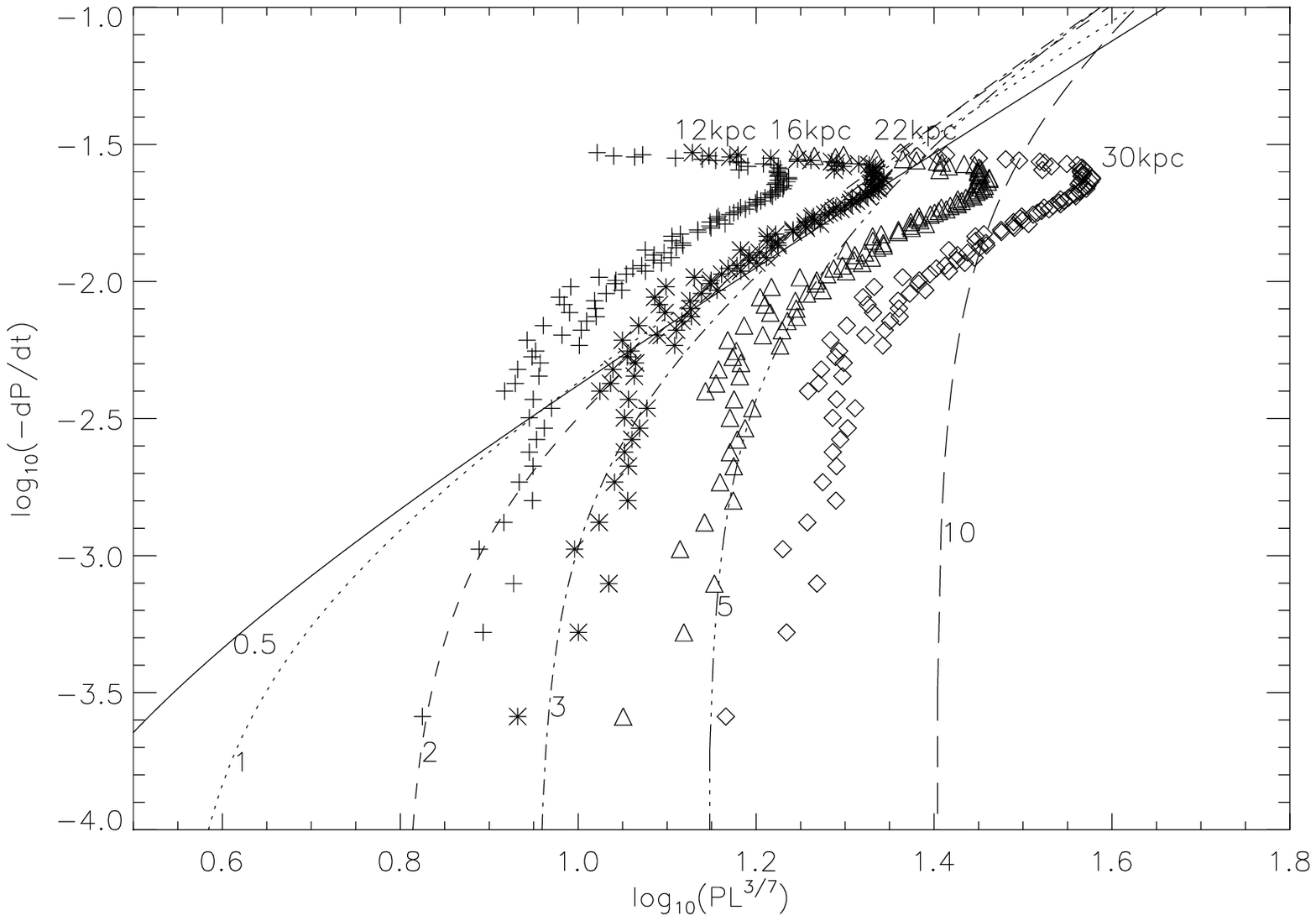}
   \caption{Plot of $\dot{P}$, taken from the spin-up model measured based on the \integral~data (section~\ref{sec:spinisg}), against the quantity $PL^{3/7}$ derived from the 15--50~keV \swiftbat~count rates scaled for various distances (points).  Each series of points is labelled with the distance, in kpc, used to estimate the luminosity from the \swiftbat~count rates.  The different lines show the model of \citet{1979ApJ...234..296G} for a 1.4~$M_{\sun}$, $10^{6}$~cm NS for various values of the neutron star magnetic moment, $\mu$ (indicated towards the bottom of each line).  The units of $\dot{P}$, $P$, $L$ and $\mu$ are s~yr$^{-1}$, s, $10^{37}$~ergs~s$^{-1}$ and $10^{30}$~G~cm$^{3}$ respectively.  The value of $\log_{10}(PL^{3/7})$ for a 4.45~s pulsar emitting at the Eddington luminosity of $\sim 2\times 10^{38}$~erg~s$^{-1}$ is approximately 1.2; a value of 1.4 is approximately 3 times $L_{Edd}$.}
              \label{fig:ghosh}%
    \end{figure}

The intrinsic spin-up model (Equation~\ref{eqn:spinup}) was then used to calculate $\dot{P}$ for the duration of each daily BAT measurement.  The results, calculated for various values of $d$, are presented in Fig.~\ref{fig:ghosh} against the models of \citet{1979ApJ...234..296G} for a 1.4~M$_{\sun}$ NS with a radius of $10^{6}$~cm and various values of $\mu$.  Several results are shown for the BAT data scaled for different distances.  

It can clearly be seen in  Fig.~\ref{fig:ghosh} that from the peak of the outburst, when $PL^{3/7}$ is highest, that the BAT data show a slope that closely matches the gradient of the model, before turning over when the magnetic torque is expected to start to brake the pulsar.  {\correction This requires confidence in the shape of the BAT light curve; a similar effect could be caused if there was an over estimation of the flux from \gro~towards the end of the outburst.  In section~\ref{sec:timingbat}, the spin-up of the pulsar was found to end at MJD~54611, which is when the 15--50~keV flux from \gro~is approximately equal to that from GX~$3+1$ (Fig.~\ref{fig:batlc}).  The separation of the two sources is 18\arcmin, slightly less than the FWHM of the BAT angular resolution of 22\arcmin~\citep{2005SSRv..120..143B}.  Assuming a Gaussian point spread function, where the FWHM is equal to $2.35\sigma$, the maximum possible contamination of a measured flux caused by an equally bright source located 18\arcmin~away is $\sim$~5\%.  If the measured flux from \gro~was over-estimated by this amount it would have a negligible effect on the appearance of the data in Fig.~\ref{fig:ghosh} - equivalent to causing a value of $\log(PL^{3/7}) = 1$ to be changed to $\sim 0.97$.  In addition, the flux from GX~$3+1$ shown in Fig.~\ref{fig:batlc} remains reasonably constant throughout the outburst and no significant correlation was found between the two light curves.  Finally, the ability of the BAT {\sevensize batcelldetect} software, which is used to prepare the transient monitor light curves, to succesfully resolve two sources is thought to be accurate where the separation is more than 10\arcmin~(Krimm, private communication).  Therefore any contamination of the \gro~flux is likely to be considerably less than 5\%}.

The value of $\mu$ which seems to most closely resemble the model is $\sim 3\times 10^{30}$~G~cm$^{3}$, corresponding to a magnetic field at the surface of the NS of $\sim 2\times 10^{12}$~G. Fig.~\ref{fig:ghosh} suggests a distance to \gro~of around 16~kpc.  However, estimating the distance from this model is uncertain due to the difficulties in estimating the spectrum of \gro~and the crude scaling of the flux values against the Crab Nebula, which is quite a different object spectrally to \gro; a variation in the bolometric correction of a factor of 2, for example, would lead to a change in the distances by a factor of $\sqrt 2$.   {\correction Nevertheless, this makes it very likely that the object is located some distance beyond the Galactic centre (approx 8.5~kpc from Earth), which is reinforced by the non-detection of any optical counterpart.  A value of 30~kpc, which would put \gro~outside the Galaxy, seems unlikely; additionally, a distance greater than 16~kpc would make \gro~the furthest HMXB yet detected by \integral~\citep{2007A&A...467..585B}.  A reasonable estimate on the range of acceptable distances measured here is therefore 12--22~kpc.  HMXBs  are relatively young systems that should be found near to their birth places in star formation regions in the spiral arms of the Galaxy;  the 4-arm spiral model of \citet{2003A&A...397..133R} gives distances to the Perseus ($\sim$~14~kpc) and Cygnus ($\sim$~16~kpc) arms that are within the range measured here.}   Note that the 18~kpc distance lower limit estimated by \citet{scott} was based on an assumed pulsed fraction of 30\%;  if the pulsed fraction during the 1995 outburst was $\sim$~15\%, as measured in section~\ref{sec:pf}, their distance would be reduced to $> 13$~kpc.  Whilst the distance estimates are uncertain, the change in $\dot{P}$ due to the accretion torque predicted by \citet{1979ApJ...234..296G} seems to agree with the data. 

The strong magnetic field estimated at  $\sim 2\times 10^{12}$~G opens the possibility to observe a Cyclotron Resonance Scattering Feature (CRSF) in the X-ray spectrum of \gro.  To date CRSFs have been detected from 16 accreting pulsars \citep{2007A&A...472..353S} and in V~0332$+$53 have also been observed to evolve with time \citep{2005A&A...433L..45K, 2006A&A...451..187M, 2006MNRAS.371...19T}.  In the case of \gro~the fundamental CRSF should be centred near to 22~keV; this would be challenging to observe with \integral~as the peak would lie near to the boundaries of the ISGRI and JEM-X sensitivities and the peak flux from \gro~is about a factor of 5 less than for V~0332$+$53, but is something that will be investigated in the future.  Similarly, the monitoring capabilities of BAT could allow very detailed measurements of the evolution of CRSFs during a long outburst, if the sensitivity in narrow spectral bins was optimised.

\begin{figure}
   \centering
   \includegraphics[width=9cm]{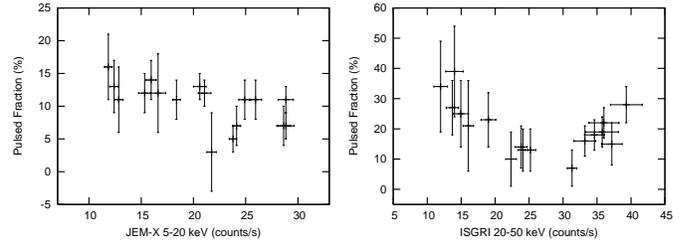}
   \caption{Plots of the \integral~ measurements of the pulsed fraction against total pulsed+unpulsed flux during the 2008 outburst of \gro.}
              \label{fig:c_pc}%
    \end{figure}

 It is interesting that in both 5--20 and 20-50~keV energy bands the pulsed fraction maxima during the early part of the outburst appear near the periastron of the binary system, and that the minima are near apastron; however, the connection between the bands is broken in \integral~revolutions 661--664 and in the following orbit no rise and fall of the pulsed fraction is visible within the error bars.  Unfortuantely, even if there where some correlation of the behaviour of \gro~with the orbital period it would be very difficult to confirm it with observations that sparcely cover only a few orbital phases.  The behaviour of the pulsed fraction with the overall brightness of the source is shown in figure~\ref{fig:c_pc}.  Above 20~keV there is a clear evolution of the pulsed fraction with flux, with opposite behaviour below and above 30~cps:  at the beginning of the outburst (right of figure~\ref{fig:c_pc}) the pulsed fraction decreases proportionally as the source dims towards 30~cps;  below 30~cps the pulsed fraction increases with decreasing flux.  Note that the first point above 30~cps is the one taken during  revolution 660, where the pulsation was at its weakest in 20--50~keV and not measurable in 5--10~keV (see figure~\ref{fig:profiles}).  This also seems to coincide with a large drop in the average 20--50~keV flux between revolutions 660 and 661.  The same effect is not visible in the lower energy band which, except for a minimum again during revolution 660, shows a slow increase in the pulsed fraction with reducing count rate.

The different behaviour above and below 20~keV rules out the attenuation of the pulsed flux by some intervening material that slowly accumulates and then dissipates; in this scenario the lower energy X-rays would show at least as strong an effect as at high energies.  Work continues to investigate these effects with fine resolution temporal-spectral analysis.

On their own, the pulse period measurements made using the BAT non-imaged count-rate data would be difficult to interpret because of the ambiguities in identifying the correct period, due to aliasing of the periodic signal, and the poor quality of the data caused by the background sources in the very large field of view.  However, once these results are compared with the very high quality pulsation measurements from ISGRI it is very easy to see that the time evolution of the pulsar can be measured over a long time scale with BAT.  This suggests that outbursts from other BeXRBs and other reasonably bright pulsars with periods $>$~few~seconds could be equally well measured and shows the power of having two complimentary instruments in \swift~and \integral~simultaneously operational.  If a proper treatment of the background in the detector plan count-rate data led to an increase in sensitivity then it may be feasible to follow the pulsations from numerous objects, for example the large numbers of BeXRBs found in the Magellanic clouds.

\section*{acknowledgements}
Based on observations with INTEGRAL, an ESA project with instruments and science data centre funded by ESA member states (especially the PI countries: Denmark, France, Germany, Italy, Switzerland, Spain), Czech Republic and Poland, and with the participation of Russia and the USA.  \swiftbat~transient monitor results provided by the \swiftbat~team; particular thanks to H. Krimm for helpful support. 2--10~keV quick-look results provided by the ASM/RXTE team.  SS's work was conducted despite support from PPARC/STFC.  JC acknowledges financial support from ESA-PRODEX 90057.  The anonymous referee is thanked for a prompt response and helpful comments.  A.J. Bird, V. McBride and T. Maccarone are also thanked for useful discussions.

\bibliographystyle{mn2e}
\bibliography{gro1750}

\label{lastpage}

\end{document}